\documentclass[10pt]{article}
\usepackage{graphicx}
\usepackage{fullpage,amsmath,amssymb,float,epic,epsfig,color}
\title{Multi-scale modeling of follicular ovulation as a reachability problem} 

\author{Nki Echenim \footnotemark[2]\ 
\and Frederique Cl\'ement \footnotemark[2]\ 
\and Michel Sorine \footnotemark[2]}
\date{}

\begin{document}
\maketitle
\renewcommand{\thefootnote}{\fnsymbol{footnote}}

\noindent \footnotetext[2]{\noindent
Unite de Recherche INRIA Rocquencourt, Domaine de Voluceau, Rocquencourt
BP~105, 78153 Le Chesnay Cedex, {\sc France}\\
tel: +33 1 39 63 57 41\\
fax: +33 1 39 63 57 86}

\renewcommand{\thefootnote}{\arabic{footnote}}

\begin{abstract}
During each ovarian cycle, only a definite number of follicles ovulate, while the 
others undergo a degeneration process called atresia. We have 
designed a multi-scale mathematical model where ovulation and atresia result
from a hormonal controlled selection process. A 2D-conservation law describes the age and maturity structuration 
of the follicular cell population. In this paper, we focus on the
operating mode of the control, through the study of the characteristics 
associated with the conservation law. We describe in particular the set of microscopic initial
conditions leading to the macroscopic phenomenon of either ovulation or atresia, in the framework of backwards reachable sets theory.\\

\end{abstract}

\section{Introduction}
\noindent The development of ovarian follicles is a crucial process for
reproduction in mammals, as its biological meaning is to free fertilizable oocyte(s) at the time of ovulation. A better understanding of follicular development is both a clinical and zootechnical challenge; it is required to
improve the control of anovulatory infertility in women, as well as ovulation rate
and ovarian cycle chronology in domestic species.\\
Within all the developing follicles, very few actually reach the ovulatory size;
most of them undergo a degeneration process, known as atresia
\cite{green_94}. The ovulation rate (number of ovulatory follicles per cycle) results
from an FSH (Follicle Stimulating Hormone)-dependent follicle selection
process. FSH acts on the cells surrounding the oocyte, the
granulosa cells, and controls both their commitment towards either
proliferation, differentiation or apoptosis and their ability to secrete
hormones such as estradiol. The whole estradiol output from the ovaries is
responsible for exerting a negative feedback on FSH release. Following the
subsequent fall in plasmatic FSH levels, most of the follicles undergo atresia and only the ovulatory ones
survive in the FSH-poor environment.\\ 
We have proposed a mathematical model, using both
multi-scale modeling and control theory concepts, to describe
the follicle selection process \cite{nous}. For each follicle, the cell
population dynamics is ruled by a conservation law with variable
coefficients, which describes the changes in age and maturity of the granulosa
cell density. A control term, representing FSH signal, intervenes both in the
velocity and loss terms of the conservation law. Two acting controls are distinguished: a global control resulting from the ovarian feedback and corresponding to FSH plasmatic levels, and a local control,
specific to each follicle, accounting for the modulation in FSH bioavaibility due to follicular vascularization. Both ovulation triggering and follicular ovulation depend on the reaching of a target.\\
In this paper, we aim at using control theory to characterize 
follicular trajectories and the control laws leading
to ovulation or atresia. The macroscopic phenomenon of follicular ovulation is considered as a reachability
problem for the microscopic characteristics associated
with the conservation law. The problem is enounced in details and
solved in open-loop. Follicles
are assumed to ovulate if their state variables, (age, maturity and cell density)
reach a given target set. Backwards reachable sets theory is used
to define the initial conditions compatible with latter ovulation, for a given control panel.
In section 2, the conservation laws describing the model for follicle selection
and their characteristics are presented. In section 3, the target corresponding to
follicular ovulation is defined. Reachable sets theory is used to solve the corresponding control
problem and simulation results are discussed. Section 4 is devoted to the discussion and perspectives.

\section{Follicle selection model}

\subsection{Controlled conservation laws for granulosa cells}
\begin{figure}[ht]
\centering 
\includegraphics[scale=0.5]{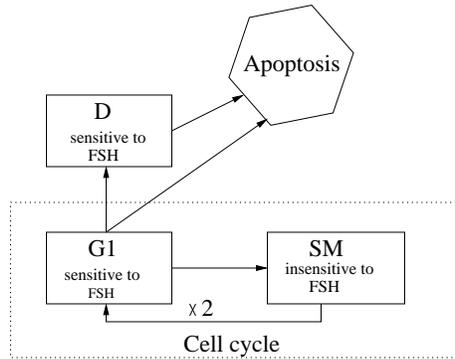}
\caption{Cell flow chart. The cell cycle consists of the cyclic G1-SM-G1 pathway. 
When it enters G1 from SM, a mother cell gives birth to two daughter cells. Cell differentiation 
corresponds to the one-way flow from G1 into D. Entry into apoptosis arises from the G1 and D phases. 
The G1 and D phases are under the control of FSH signal.}
\label{cycle}
\end{figure}

Following  \cite{nous},  cells are characterized by their positions within or outside the cell cycle and by their sensitivities to FSH. 
This leads to distinguish 3 cellular phases within the granulosa cell population. Phases~1 and 2 correspond to the proliferation phases (describing respectively the G1 phase and S to M phases of the cell cycle), and phase~3 corresponds to the differentiation phase, after cells have exited the cell cycle (see Figure \ref{cycle}). 

\begin{figure}[ht] 
\centering 
 \includegraphics[width=8cm]{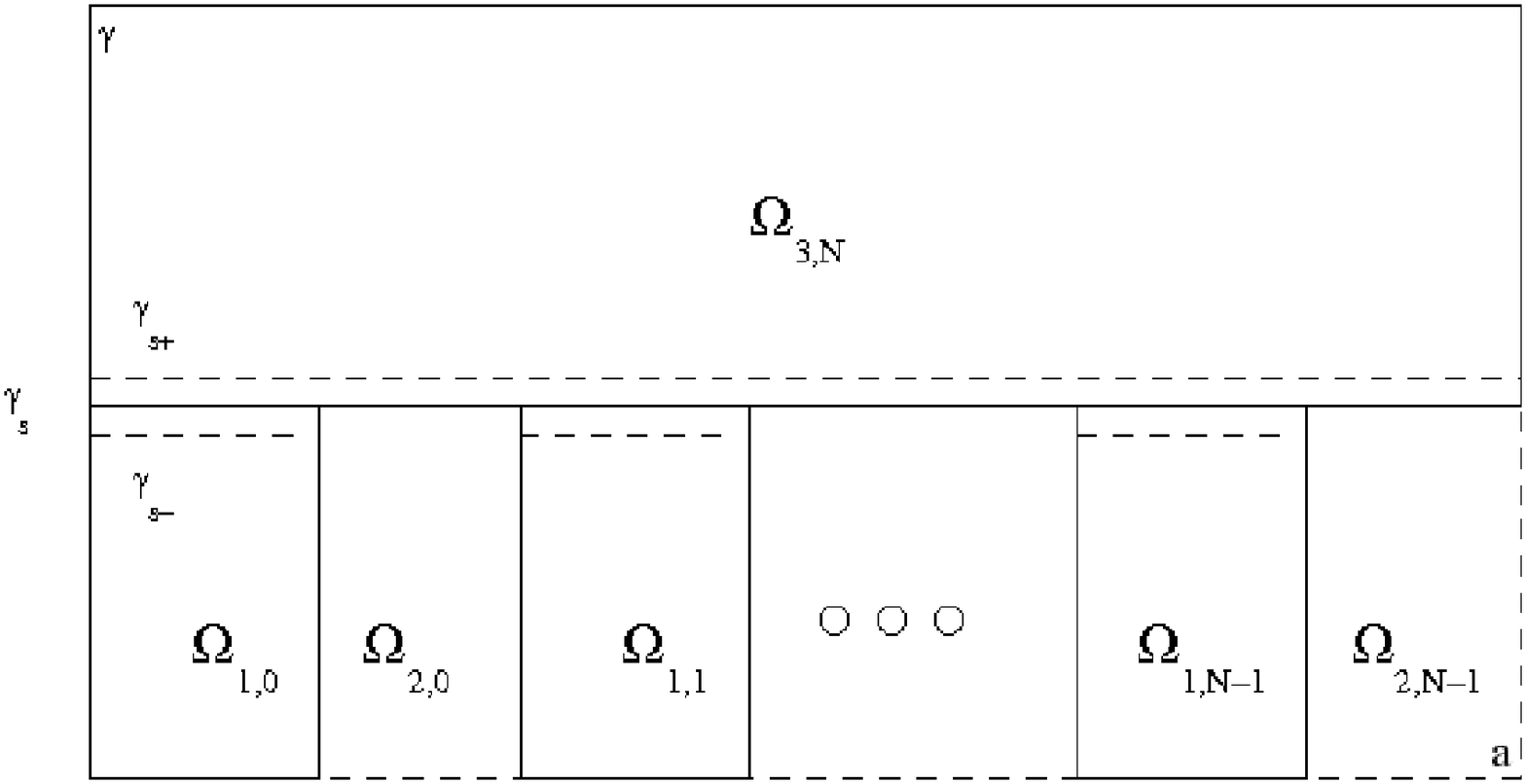}  
\caption{Cellular phases on the age-maturity plane. The cell cycle is represented by the pathway $\Omega_{1,k}$, $\Omega_{2,k}$... and 
the differentiation phase by $\Omega_{3,N}$. }
\label{dom}
\end{figure}

\noindent More precisely, the position of a cell at time $t \in [0, T]$ ($T > 0$) is defined by its age and maturity. 
The cell age $a$ is a marker of progression within the cell cycle (in phases 1 and 2) and evolves as time $t$ outside the cycle (in phase 3). The duration of phase 1 is $a_1 >0$ and the total cycle duration is $a_2$ (so that phase 2 duration is $a_2 -a_1 >0$). 
The maturity marker $\gamma $ is used to sort the cycling and non-cycling cells by comparison to a threshold $\gamma_s$ and to characterize the cell vulnerability towards apoptosis (programmed cell death). 
Phases 1, 2, 3 correspond respectively to ranges in the values of $(a, \gamma )$ in the following open sets of the age-maturity plane: (see Figure \ref{dom}): 
$$ 
\Omega_{1,k} =  ]ka_2, k a_2+a_1[ \times ]0, \gamma_s[,  \quad 
\Omega_{2,k} =  ]ka_2+a_1, (k+1) a_2[ \times ]0, \gamma_s[, \quad 
\Omega_{3,k} =  ]0, (k+1) a_2 [ \times ]\gamma_s, \gamma_{max}[
$$ 
for integers $k=0, 1, \ldots $. We will also use the notation $Q_{j,k}=\Omega_{j,k} \times ]0,T[ $ for $j=1,2,3$. 

\noindent The cell population in a follicle, $f$, is represented by cell density functions, $\phi^k_{f,j}(a,\gamma ,t)$, defined on 
each cellular phase $Q_{j,k}$, $j=1,2,3$, $k=0,1, \ldots $ as solutions of the following conservation laws  \cite{nous}: 
\begin{eqnarray}
\label{cons}
\frac{\partial \phi^k_{f,j}}{\partial t}&+&\frac{\partial g_{f}(u_f)\phi^k_{f,j}}{\partial a} + 
\frac{\partial h_{f}(\gamma , u_f)\phi^k_{f,j}}{\partial \gamma} = 
-\lambda(\gamma,U)\phi^k_{f,j} \quad \mbox{in } Q_{j,k}, \; j=1,2,3, \; k=0,1, \ldots  
\end{eqnarray}  

\noindent In phases 1 and 3, both a global control, $U$, and a local control, $u_f$, act on the velocities of aging 
($g_{f}$ function, locally controlled in phase 1) and maturation ($h_{f}$ function, locally controlled in phases 1 and 3) 
as well as on the loss term ($\lambda$ apoptosis rate, globally controlled in phases 1 and 3). 
Phase 2 is uncontrolled and corresponds to completion of mitosis after a pure delay in age, $a_2 -a_1$ (no cell can leave the cycle here).  
We have the following expressions for those functions, for  $k=0,1, \ldots$ (all parameters are real positive numbers defined in Table~\ref{param}): 

 \begin{eqnarray} 
\label{vitesseg13} g_{f}(\gamma,u_f) &=& \tau_{gf}(1 - g_1\omega_{s_-}(\gamma) (1 - u_f)) \quad \mbox{in } \Omega_{j,k}, \; j=1, 3  \\ 
\label{vitesseh13}  h_f(\gamma,u_f)&=&\tau_{hf}(-\gamma ^2 + (c_1\gamma+c_2)(1- \omega_{s}(\gamma) \exp(-u_f/\bar{u}))) 
 \quad \mbox{in } \Omega_{j,k}, \; j=1, 3 \\
\label{vitessel13}  \lambda(\gamma,U)&=&\omega_{\lambda}(\gamma)(1-U) \quad \mbox{in } \Omega_{j,k}, \; j=1, 3 \\   
 \label{vitessegh2} g_{f}(u_f)&=&\tau_{gf}, \; \; h_f(\gamma,u_f) =  \lambda(\gamma,U) = 0  \quad \mbox{in } \Omega_{2,k}
\end{eqnarray}
The $\omega_{s_-}$ and $\omega_{s_+}$ functions are smooth  switches between $0$ and $1$ in a strip defined for all $a$, 
$0 < \gamma_{s_-} < \gamma < \gamma_{s_+}$ containing $\gamma_{s}$  
(we use the notations $|x|_+ = \max (x,0)$ and $|x|_- = \max (-x,0)$):  
$$ 
\omega_{s_-}(\gamma)= \exp \left( -\frac{|\gamma - \gamma_{s_-}|_+}{|\gamma - \gamma_{s}|}\right), \; 
\omega_{s_+}(\gamma)= \exp \left( -\frac{|\gamma - \gamma_{s_+}|_-}{|\gamma - \gamma_{s}|}\right), \; 
\omega_{s}(\gamma) = \omega_{s_-}(\gamma) + \omega_{s_+}(\gamma). 
$$ 
The factor $\omega_{s_-}$ in (\ref{vitesseg13}), switches off the control of $g_f$ in phase~3Ê: 
$g_f = \tau_{gf}$ in $\Omega_{3,k}$,  and switches it on for $(a, \gamma)$ strictly in phase~Ê1, where    
$g_{f}(u_f) = \tau_{gf}(1 - g_1 + g_1 u_f))$ for $\gamma \leq \gamma_{s_-}$.  \\ 
In the same manner, $\omega_{s}$ in (\ref{vitesseh13}), switches off the control of $h_f$ for $\gamma = \gamma_{s}$, so that, 
in a neighbourhood  of $\gamma_{s}$, $h_f > 0$ for all $t$ and $u_f$: the boundary $\gamma = \gamma_s$ is inward for 
$\Omega_{3,\infty}$ 
and outward for each $\Omega_{1,k}$. 
In particular, a cell very close to exit the cycle (in phase~1) is committed to go into phase~Ê3. The exiting flux is controlled by $u_f$ 
inside phase~1 before the cell maturity reaches $\gamma_s$: the sign of $h_f$ can be changed by the control only 
below $\gamma_{s_-}$. \\ 
In complement to this local control (by $u_f$) of the exit flux from the cell cycle, a global control is exerted (by $U$, $ U\leq 1$) on the cell vulnerability towards apoptosis, in a zone surrounding $\gamma_{s}$, through the function 
$\omega_{\lambda}(\gamma)=K\exp\left(-\left(\frac{\gamma - \gamma_s}{\bar{\gamma}}\right)^2\right)$ in (\ref{vitessel13}).

\noindent Each boundary segment of each (rectangular) $\Omega_{j,k}$, $j=1,3$,  $k=0,1, \ldots $ is either inward or outward, independently on time $t$ or controls $u_f$, $U$, so that we can consider the following boundary conditions (the use of the trace values and the wellposedness of this model will be discussed later): 

\medskip 

\noindent Boundary conditions for $\phi^k_{f,1}$, $t \in ]0,T[$ :  
\begin{eqnarray}
\label{BC1a}   g_{f}(\gamma,u_f) \phi^k_{f,1}(ka_2,\gamma,t) &=&  \left\{ \begin{array}{l} 
2\tau_{gf}\phi^{k-1}_{f,1}(ka_2,\gamma,t), \;  \mbox{for }   k \geq 1 ,   \\ 0, \;  \mbox{for } k=0  
 \end{array}\right.   \;  \mbox{for } \gamma \in ]0, \gamma_s[. \\
\label{BC1g}  \phi^k_{f,1}(a,0,t) &=& 0, \; \mbox{for } a \in ]ka_2, ka_2+a_1[ 
\end{eqnarray} 
\noindent Boundary conditions for $\phi^N_{f,3}$, $N \geq 1$, $t \in ]0,T[$ :  
\begin{eqnarray} 
\label{BC3a}   \phi^N_{f,3}(0,\gamma,t) &=&  0, \;  \mbox{for } \gamma \in ]\gamma_s, \gamma_{max}[  \\ 
\label{BC3g}   \phi^N_{f,3}(a,\gamma_s,t) &=& \left\{ \begin{array}{l} 
 \phi^k_{f,1}(a,\gamma_s,t), \;  \mbox{for } a \in [ka_2, ka_2+a_1[ \\ 0, \;  \mbox{for } a \in [ka_2+a_1, (k+1)a_2[  
 \end{array}\right. , \; \mbox{for } k= 0, \ldots N-1. 
\end{eqnarray}

\noindent Initial conditions:
\begin{equation}
\label{initialisation} 
\phi^k_{f,j}(a,\gamma,0) =\phi^k_{f0}(a, \gamma)|_{\Omega_{j,k}}, \; j=1,3. 
\end{equation}

\noindent In (\ref{BC1a}), the delay $a_2 - a_1$ and the doubling of the flux due to mitosis have been explicitly taken into account, so that it is not necessary anymore to consider  phase~2 and $\phi^k_{f,2}$ in the model. The cell density of interest in a follicle, $\phi^N_f$, is simply:  
\begin{equation*}
\phi^N_f=\phi^k_{f,1} \mbox{ on }Q_{1,k},\,  \; k = 0, \ldots N-1, \; \phi^N_f=\phi^N_{f,3} \mbox{ on }Q_{3,N}
\end{equation*} 
\noindent In practice, we will fix the value of $N$ to a large enough integer, and use the notations $\phi_f$ and 
$a_{max} = Na_2$.  $\phi_f$ can be computed by solving the series of problems:  
\begin{eqnarray} 
\label{Pb1} 1. & \; P_{1,k} : & \mbox{ (\ref{cons}), (\ref{BC1a}), (\ref{BC1g}), (\ref{initialisation}) on } Q_{1,k}, \; 
\mbox{ for } k= 0, \ldots N-1. \\ 
\label{Pb3} 2. & \; P_{3,N}:  & \mbox{ (\ref{cons}), (\ref{BC3a}), (\ref{BC3g}), (\ref{initialisation}) on }Q_{3,N}, \; \mbox{when the } P_{1,k}  \; 
\mbox{are solved }. 
\end{eqnarray}

\noindent It is worth noticing that those problems can be solved in sequence when the traces of the solutions of $P_{1,k}$ on their outward boundaries (left and upper segments) are well defined. This is a useful property for the efficiency of computations and the backward reachability technique. 

\begin{table}[ht]
\centering
\begin{tabular}{cllc}
\hline
\tt Functions and parameters & Definition & Nominal value\\
\hline
\tt $S[M(\sum_f{\phi_f})]$ &Global feedback control\\
\tt $U_{s}$& minimal plasmatic FSH value  & $0.5$\\
\tt $c$ & slope parameter of the sigmoid function &$0.1$\\
\tt $m$  & abscissa of the inflexion point of the sigmoid function &$50$ \\
\tt $\tau$  & FSH delay from the pituitary gland to the ovaries & $0.01$ \\
\hline
\tt $b_f[M(\phi_f)]$ & Local feedback gain\\
\tt $b_1 $ &basal level & $0.054$\\
\tt $b_2$ & exponential rate& $ 0.3$ \\
\tt $b_3$ & scaling parameter & $27$\\
\tt $\tau_f$  & delay to modify local vascularization & $0.01$ \\
\hline
\tt $g_f( u_f)$ &Aging velocity\\
\tt $\tau_{gf}$  & time scale parameter&  1\\
\tt $g_1$ & control gain in phase 1 &  0.5\\ 
\hline
\tt $h_f(\gamma, u_f)$&Maturation velocity\\
\tt $\tau_{hf}$ & time scale parameter& 0.07 \\
\tt $c_1$ & slope parameter& 11.892\\
\tt $c_2$& origin ordinate& 2.288\\
\tt $\bar{u}$ & scaling parameter& 0.133\\
\hline
\tt $\omega_{\lambda}(\gamma)$&Global feedback gain\\
\tt $K$ &amplification constant& 3 \\
\tt $\bar{\gamma}$ & scaling parameter& 0.2\\
\hline
$a_1$ & cellular age at the end of phase 1& 1\\
$a_2$ & cellular age at the end of phase 2& 2\\ 
$N$ & maximum number of cell cycles & 8 \\ 
\hline
$\gamma_s$ & maturity threshold for cell cycle exit& 3\\
$\gamma_{s_-}$ & maturity threshold for switching off control in phase~1& 2.99 \\
$\gamma_{s_+}$ & maturity threshold for switching on control in phase~3 & 3.01 \\
$\gamma_{max}$ & maximum maturity  & 15 \\
\hline
$M_s$ &  ovarian threshold for ovulation triggering& 75 \\
$M_{s1}$ &  follicular threshold for ovulation ability & 40 \\ 
\hline
\end{tabular}
\caption{Main model functions and parameters}
\label{param}
\end{table}

\begin{figure}[t]
\begin{center}
\includegraphics[width=7cm]{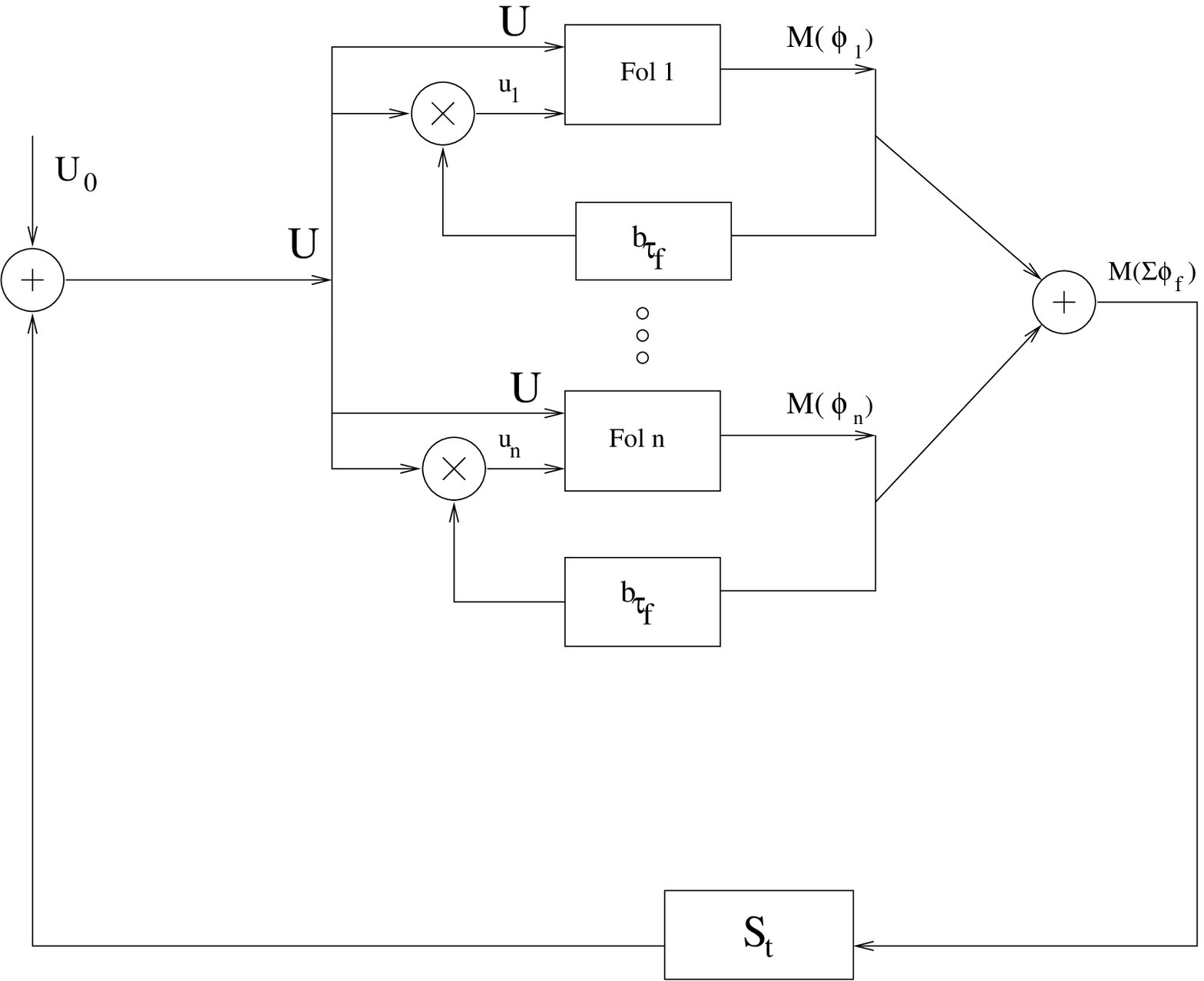}
\caption{Ovarian system: the multi-scale model in closed-loop. {\it The global feedback
control $U$ is modulated on the follicle scale ($Fol~i$), by the follicular maturity, $M(\phi_i)$, to define the local control $u_{fi}$.}}
\label{closed-loop}
\end{center}
\end{figure}

\noindent The feedback exerted by the ovaries on the secretion of the hormonal control
FSH defines a closed-loop system (cf. Figure \ref{closed-loop}). The global
control $U$ can be interpreted as FSH plasmatic level. The local control
$u_f$ represents intra-follicular bioavailable FSH levels. It is given as a proportion of $U$. \\
Define the maturity operator $M$ as: 
\begin{equation}
\label{moment}
M(\varphi)(t) = \int_{0}^{\gamma_{max}}\!\int_0^{a_{max}}{\gamma\varphi(a,\gamma,t)dad\gamma}
\end{equation}
The global maturities $M(\phi_f)$ on the follicular scale, and $\displaystyle M(\sum_f{\phi_f})$ on the ovarian scale, 
will be used to define the two-scale feedback control: 

\begin{eqnarray}
\label{uf}
\nonumber U &=& S_{\tau}(M(\sum_f{\phi_f}))+U_0 \\ 
u_f & = & b_{f \tau_f}(M(\phi_f)).U \\
\nonumber \mbox{ where } S(\mu) &=& U_{s}+\frac{1-U_s}{1+\exp\left(c(\mu - m)\right)}, \quad  
S_{\tau}(\mu)(t) = S(\mu(t -\tau )) \\ 
\nonumber b_{f }(\mu) &=& b_1 +\frac{1-b_1}{1+\exp(b_2(b_3 - \mu))}, \quad  
b_{f \tau_f}(\mu)(t) = b_{f }(\mu(t - \tau_f))
\end{eqnarray}

\noindent The decreasing sigmoid function $S$ accounts for the ovarian negative feedback on FSH~Ê; 
$U_0(t)$ is a potential exogenous entry. 
The increasing sigmoid function $b_f$ makes $u_f$ increase with the maturity of follicle $f$, up to its reaching FSH plasmatic values. 
The delay $\tau$  is introduced to take into account the time needed for FSH signal to reach the ovaries; and the local delay $\tau_f$, that can be neglected in practice ($\tau_f \ll \tau$) is mainly here to facilitate the resolution on finite time intervals. 
The delayed versions of $S$ and $b_{f }$ are denoted $S_{\tau}$ and $b_{f \tau_f}$ respectively. 
The parameters are chosen such that $U$ and $b_f$ are normalized: $0 < b_1, \; U_s < 1$, so that: 
 \begin{equation}
\label{normal-u}
0 \leq u_f \leq U \leq 1 + U_0. 
\end{equation} 

\noindent Ovulation is triggered when estradiol levels reach a threshold value $M_s$. As estradiol secretion is related to
maturity (see \cite{nous}), the ovulation time $T_s$ is defined as: 
\begin{equation}
\label{mmt-ovarien}
T_s = \min \{T \; | \; M(\sum_f{\phi_f})(T) = M_s \} 
\end{equation}
The follicles are then sorted according to their individual
maturity. The ovulatory follicles are those whose maturity at time $T$ has overpassed a threshold
$M_{s1}$ such as $M_{s1}\leq M_s$.
The ovulation rate is computed as: 
\begin{equation} 
N_{s,s_1} = \operatorname{Card}\{f \; | \; M(\phi_f)(T_s)\geq M_{s1}\}
\end{equation}

\begin{figure}[t]
\begin{center}
\includegraphics[width=15cm]{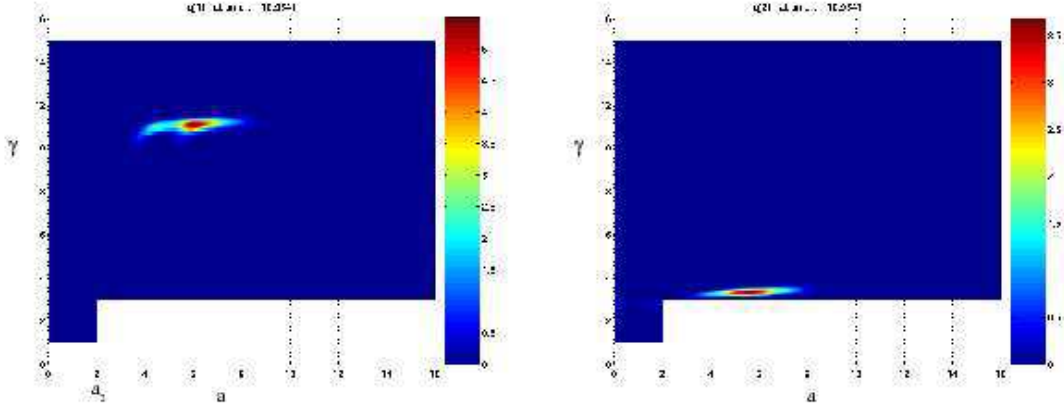}  
\caption{Cell density of an ovulatory follicle (left) and an atretic follicle (right). The horizontal axis represents the cell age, the
vertical axis the cell maturity, and the colorbar indicates the cell density value (number of cells per elementary age $\times$ maturity volume). $a_2$ is the cell cycle duration (2 age units).}
\label{repartition}
\end{center}
\end{figure}

\noindent For instance, Figure \ref{repartition} shows, on the age-maturity domain, the cell density of either an ovulatory follicle (left panel) or an atretic one (right panel) 
at ovulation time. The cell cycle is implemented on a periodic domain, where the age $a$ is reset after the cells
go through mitosis. The granulosa cell repartition in the ovulatory follicle is characterized by a roughly older age range, and a
higher maturity and cell density ranges than in the atretic follicle.

\subsection{Wellposedness of the model}
We have obtained a compact expression of the model by using a density representation 
of the cell population instead of a discrete model with a huge number of cells. Even if reasoning about 
controlled trajectories of individual cells will be useful in the sequel, the global numerical simulations 
are based on this compact form. It is then useful to discuss the wellposedness of the model in order to 
check, to some extent, its consistency.  \\
Due to the delays in the feedback control laws, we can consider the $u_f$ and $U$ control terms  as given functions of time on each interval of size $\tau_f$. The velocities and loss term are then given 
functions  $g_f(\gamma,t)$, $h_f(\gamma,t)$ and $\lambda(\gamma,t)$. In this case, we show below that there 
exists a unique solution to the series of problems (\ref{Pb1}), (\ref{Pb3}).\\
In each domain $Q_{1,k}$, $Q_{3,N}$, the velocity field $(g_f, h_f)$ and the loss rate $\lambda$ can be 
considered, without loss of expressiveness for the model, as smooth functions. Yet, initial and boundary 
values may be irregular, so that we need to check the existence of weak solutions. \\ 
As mentioned below, it is sufficient to solve the sequence of $P_{1,k}$ and then $P_{3,N}$ to define $\phi_f$. 
Since those problems have a common structure, we just need to study the solution of $P_{1,k}$ and its
 traces on the outward boundary used to define both $P_{1,k+1}$ and $P_{3,N}$. 

\medskip 

\noindent We first use a change of variables to transform (\ref{cons}) into a conservation
equation without a loss term. 
Let $\phi^k_{f,1} = \tilde{\phi^k_{f,1}}\exp{(l(\gamma,t))}$ defined on $Q_{1,k}$. 
Let $\partial Q_{1,k}^+$ and $\partial Q_{1,k}^-$ denote the inward and outward boundaries of the domain $Q_{1,k}$, and let $l(\gamma,t)$ such as
\begin{eqnarray}
\label{l}
\frac{\partial
l(\gamma,t)}{\partial t} +h_f(\gamma,t)\frac{\partial
l(\gamma,t)}{\partial \gamma}&=&-\lambda(\gamma,t), \quad l(\gamma,0) = 0
\end{eqnarray} 
Here, we can assume that $h_f$ and $\lambda$ (extended to all $\gamma \in \mathbb{R}$) are $C^1$ functions, so that, according to standard theory, there exists a unique generalized solution $l$ of the linear transport equation (\ref{l}), that is Lipschitz for all $t>0$. 
Now $\tilde{\phi^k_{f,1}}$ verifies
\begin{equation}
\label{phitilde}
\frac{\partial \tilde{\phi^k_{f,1}}}{\partial t}+div{(v_f\tilde{\phi^k_{f,1}})}=0
\end{equation}
where $v_f$ is the (smooth) velocity vector: $v_f=\left(
\begin{array}{c}
g_f(\gamma,t)\\
h_f(\gamma,t)\\
\end{array}
\right).
$ 

\noindent Non homogeneous boundary conditions on $\partial Q_{1,k}^+$, similar to (\ref{BC1a}), (\ref{BC1g}) complete (\ref{phitilde}). 
To solve such a problem, with $L^{\infty}$ or $L^{1}$ boundary and initial values, extensions to the classical Kruzkov entropy solutions (\cite{kruzkov}) to boundary value problems have been proposed (see e.g. \cite{szepessy} and \cite{Porretta} where renormalized entropy solutions are defined for this type of problem). An additional problem faced here consists in defining the trace of the solution on the outward boundary $\partial Q_{1,k}^-$, in order to solve the next problems $P_{1,k+1}$. 
It is not clear whether this trace can be defined here for such weak solutions. As $v_f \in (L^{\infty}(Q_{1,k}))^2$ 
and $\text{div}(v_f)\in L^{\infty}(Q_{1,k})$, an alternative is to use the results of \cite{pousin}, to define a
unique solution $\tilde{\phi^k_f} \in L^2(Q_{1,k})$ with well defined trace in $L^2(\partial Q_{1,k}^-)$ as soon as the boundary data are in $L^2(\partial Q_{1,k}^+)$. A drawback of this alternative is that, in our case, it is not known whether those solutions coincide with classical weak solutions. 

\subsection{Characteristic equations}
\noindent The model deals with controlled partial differential
equations. Up to now, no convincing control strategy for such velocity controlled conservation laws with integro-differential control terms is
available from
the literature.
To tackle the control problems, we focus on the actions of the
control terms on the characteristics of the conservation laws. As the solution of the conservation laws is $L^2$, we can derive the characteristic equations.\\
We now deal with the following ordinary differential equations \cite{evans}:
\begin{eqnarray}
\label{caracteristiques}
\left\{
\begin{array}{l}
\dot{a_c}=g_f(\gamma_c, u_f)\\
\dot{\gamma_c}=h_f(\gamma_c,u_f)\\
\dot{\phi_{fc}}=-\left(\lambda(\gamma_c,U)+\frac{\partial h_f(\gamma_c,u_f)}{\partial \gamma_c}\right)\phi_{fc}\\
\end{array}
\right. \quad c=1 \dots n
\end{eqnarray}
where $g_f$, $h_f$ and $\lambda$ are defined in (\ref{vitesseg13}),(\ref{vitesseh13}),(\ref{vitessel13}),(\ref{vitessegh2}) and $n$ is the number of characteristics.\\
The local control $u_f$ acts on the position of the characteristics on the
$(a_c,\gamma_c)$ spatial domain at a given time, and $U$ on the loss term $\lambda$. 

\medskip 

\noindent The transfer conditions between the cellular phases
are continuous on $a_c$ and $\gamma_c$ and discontinuous on $\phi_{fc}$. 
From phase 1 to phase 2 (crossing the right boundary of some $\Omega_{1,k}$), the flux continuity condition reads: \\ 
- for $T_k$ such as: $a_c(T_k)=a_1+ka_2$, $0\leq\gamma_c(T_i)\leq\gamma_s$, for some $k \in \mathbb N$,  
\begin{eqnarray}
\label{c1}
\phi_{fc}(T_k^+)& = & \left(1 - g_1\omega_{s_-}(\gamma) (1 - u_f(T_k^-))\right) \phi_{fc}(T_k^-)
\end{eqnarray}
\noindent
From phase 2 to phase 1 (crossing the left boundary of some $\Omega_{1,k}$), after mitosis, the flux doubling condition reads: \\
- for $T_k$ such as: $a_c(T_k)=ka_2$, $0\leq\gamma_c(T_i)\leq\gamma_s$ for some $k \in \mathbb N$,   
\begin{eqnarray}
\label{c2}
\left(1 - g_1\omega_{s_-}(\gamma) (1 - u_f(T_k^+))\right) \phi_{fc}(T_k^+)&=&2\phi_{fc}(T_k^-)
\end{eqnarray}

\noindent Ovulatory follicles are distinguished from the atretic ones by their characteristic curves
reaching a well defined zone of the $(a_c,\gamma_c,\phi_{fc})$ space at a given time, as we can see on
Figure \ref{repartition}. In the next
section, we study to which extent characteristics curves can reach the target sets when subject to appropriate open-loop control laws. 

\section{Backwards reachable sets}\label{brs}

\subsection{Link with Hamilton-Jacobi-Bellman equations}
\noindent We assume that a follicle
ovulates (respectively undergoes atresia) at time $T$ if:\\
$\forall c=1 \dots
n,~(a_c(T),\gamma_c(T),\phi_{fc}(T)) \in \mathcal{M}_o (\mbox{ resp. }\in \mathcal{M}_a)$ where 
$\mathcal{M}_o$ and $\mathcal{M}_a$ are the targets for respectively ovulation and atresia. \\ 
We focus here on the backwards problem, and we first introduce the elements of its solution, 
the solvability tube  \cite{Kur:01} (or the backward reachable set \cite{tomlin2}) associated with $\mathcal{M}_o$ (resp. $\mathcal{M}_a$), i.e. the states from which it is possible
to reach $\mathcal{M}_o$ (resp. $\mathcal{M}_a$), given admissible controls $U$ and $u_f$. 
From a physiological viewpoint, it corresponds to the initial conditions compatible with ovulation (resp. atresia) and subject to correct control laws amongst the set of admissible controls. 

\medskip 

\noindent We recall briefly how such backwards reachability problems can be solved in the framework of optimal
control theory. Let us consider the following nonlinear continuous controlled dynamics, where  $f$ is a Lipschitz continuous function, 
so that there is a unique continuous solution for any measurable $u(t)$: 
\begin{eqnarray}
\label{syst}
\left\{
\begin{array}{c}
\dot{x}=f(t,x,u), \quad t \geq \tau \\
x(\tau)  \in \mathbb R^n\\
u \in \mathcal{U} \subset \mathbb R^m\\
\end{array}
\right.
\end{eqnarray}

\noindent Given a closed target set $\mathcal{M}\in R^n$ and two times $\tau$ and $t_1$, the solvability set 
$W(\tau,t_1,\mathcal{M})$ is the set of states $x \in R^n$ such that there exist control functions $u$ 
with values in the compact set $\mathcal{U}$ that steer system (\ref{syst}) from the state $x(\tau)=x$ to $x(t_1) \in \mathcal{M}$ \cite{Kur:01}. 
$W(\tau,t_1,\mathcal{M})$ can be computed by solving the following optimization problem 
where $d^2(x,X)=\min\{ ||x-z||^2 \; |\; z \in X\}$ is the square of the distance $d(x,X)$ from point $x$ to set $X$: 
\begin{equation}
\label{minu}
\min_{u \in \mathcal{U}}\{d^2(x(t_1),\mathcal M)|x(\tau)=x\}
\end{equation} 
The solution of this optimal control problem can be found by solving the following backward Hamilton-Jacobi-Bellman (HJB) equations, where $H_{f}(D_xV,x) = \displaystyle \min_{u \in \mathcal{U}} \{(D_xV)^T.f (t,x,u)\} $ is the Hamiltonian \cite{brysonetho}:
\begin{eqnarray}
\label{hamilton}
V_t + H_{f}(D_xV,x) = 0, \quad t \leq t_1, \quad V(t_1,x) = d^2(x(t_1),\mathcal{M}) 
\end{eqnarray} 
Then the following property is true \cite{Kur:01}:
\begin{equation}
\label{W}
W(\tau,t_1,\mathcal M)=\{x| V(\tau,x) \leq 0\}
\end{equation} 

\noindent The solvability tube is then the set-valued function of the starting time $t$ for given
$t_1$ and $\mathcal{M}$:
\begin{equation}\label{solvability}
\mathcal{W}[t_1,\mathcal{M}]: \quad t \rightarrow \mathcal{W}[t_1,\mathcal{M}](t) = W(t,t_1,\mathcal{M}), \quad \tau \leq t \leq t_1. 
\end{equation} 

\noindent Any initial time $t$ is thus associated with a delay of exactly $\left(t_1-t\right)$ to reach the target.

\noindent In contrast, the backward reachable set is the set of states from which it is possible to reach the target in at most $\left(t_1-t\right)$ \cite{tomlin2}: 
\begin{equation}\label{reachability}
\mathcal{G}(t ~; t_1,\mathcal{M}) = \displaystyle \bigcup_{t \leq s \leq t_1} W(t , s ,\mathcal{M}), 
\quad t \leq t_1. 
\end{equation} 
This set can also be  characterized in  the same manner as in expression (\ref{W}). It is the zero sublevel set of the $V$ function solution of the HJB equation:
\begin{eqnarray}\label{V-intro}
V_t + \min (0, H_{f}(D_xV,x)) &=& 0, \quad t \leq t_1, \quad V(t_1,x) = d^2(x(t_1),\mathcal{M}) \\ 
\label{G-intro} 
\mathcal{G}(t ~; t_1,\mathcal{M}) &=& \{x \in R^n \; | V(t,x) \leq 0\} 
\end{eqnarray} 

\noindent In the next section, we use the numerical methods based on equations (\ref{V-intro}), (\ref{G-intro}) to compute the $\mathcal{G}$ set \cite{Mitchell:05, toolboxmitch:04}. This choice is convenient since it seems easier to compute the $\mathcal{G}$ set than the $\mathcal{W}$ set-valued function, but it amounts to deal with desynchronized final times.
Indeed, in the $\mathcal{W}[t_1,\mathcal{M}]$ tube, every states considered at the same starting time reach the $\mathcal{M}$ target at the same final time $t_1$, whereas in $\mathcal{G} (t ~; t_1,\mathcal{M}) $ states reaching the $\mathcal{M}$ target at different final times (provided that they are not later than $t_1$) are fused. To ensure that every states reaching the target remain inside until $t_1$, the control function has to be extended.

\subsection{Application to granulosa cells} 

\subsubsection{The follicle as a controlled dynamical system}\label{exist} 
\noindent We represent each follicle by $n$ cells following characteristic curves of (\ref{cons}) 
defined by three state variables: age, maturity and cell density as in (\ref{caracteristiques}). 
The model of the follicle is then of the form (\ref{syst}) with $3n$ state variables: 
$$x(t) = (a_1(t),...,a_n(t);\gamma_1(t),...,\gamma_n(t);\phi_{f1}(t),...,\phi_{fn}(t))^T$$
\noindent and for each cellular phase $i=1,2,3$, the dynamics is defined by the following $f_i$: 

\begin{eqnarray}
\label{f1}
f_i=\left(
\begin{array}{c}
g_f(\gamma_1, u_f)\\
\vdots\\
g_f(\gamma_n, u_f)\\
h_f(\gamma_1, u_f)\\
\vdots\\
h_f(\gamma_n,u_f)\\
K(\gamma_1,u_f,U)\phi_{f1}\\
\vdots\\
K(\gamma_n,u_f,U)\phi_{fn}
\end{array}
\right)  
\end{eqnarray}
where $g_f$ and $h_f$ are defined in (\ref{vitesseg13}), (\ref{vitesseh13}), (\ref{vitessegh2}) and $K(\gamma_c,u_f,U)$ is defined by
\begin{eqnarray}
\text{Phases 1 and 3:} && \qquad \forall k, \; N  \in \mathbb{N}, \quad \forall (a_c,\gamma_c) \in \Omega_{1,k} \cup \Omega_{3,N} \nonumber \\ 
&&  \qquad  K(\gamma_c,u_f,U) = -\left(\lambda(\gamma_c,U)+\frac{\partial h_f(\gamma_c,u_f)}{\partial \gamma_c}\right) \label{expressionK} \\
\text{Phase 2:} && \qquad \forall k \in \mathbb{N}, \quad \forall (a_c,\gamma_c) \in \Omega_{2,k}  \nonumber \\
 &&  \qquad  K(\gamma_c,u_f,U) = \ln(2)\delta(a_c)\phi_{fc}, \text{ with } \int_{ka_2 + a_1}^{(k+1)a_2} \delta(a)da = 1 \label{f3}
\end{eqnarray} 

\noindent The expression of $K(\gamma_c,u_f,U)$ in phase 2 results from the regularization by a smooth positive function $\delta(a)$ of the mitosis at the transition (\ref{c2})  between phases 2 and 1, that will be used in the numerical treatment of the equations where the transmission boundaries are included in the computational domain. The precise choice for $\delta(a)$ does not matter since its only role is to guarantee that the density of cells entering phase 2 at time  $k a_1$ has doubled at time $k a_1 +a_2$. Since $a_2 - a_1 =1$, the simplest choice is $\delta(a)\!=\!1$  so that we have $\delta(a_c) \!=\! 1$, resulting in $K(\gamma_c,u_f,U) = \ln(2)\phi_{fc}$. 

\noindent According to physiological knowledge, we can consider that there exists a maximal age, $a_{max}$, beyond which
all cells become senescent and die. This assumption introduces the notion of a maximal
life time $T_{max}$ such that
$a(T_{max})=a_{max}$ ($T_{max}$ exists because $\dot{a} > g_2 > 0$). Hence, there is
also a maximal maturity $\gamma_{max}$ and a maximal cell density $\phi_{fmax}$
attainable for $t\in [0,T_{max}]$. The $f_i$ are Lipschitz continuous, so that there is a unique solution $x$ within each phase. 

\medskip 

\noindent We now state some properties of this controlled system and the corresponding reachable targets in the age-maturity plane.  We first describe the control action on the orientation of the velocity field: 
\begin{eqnarray}  
&& \text{In the interior of phases 1 \& 3:}  \quad \forall u_f  \in \mathcal{U}, \nonumber \\ 
&&  g_h \geq g_1\tau_{gf}\equiv 0.5 \tau_{gf}  \label{prop-gh} \\ 
&&  
\left\{
\begin{array}{ll}
h_f(\gamma, u_f) \geq 0, \quad  \text{for } u_f \geq u_f^*   \text{ and }  & 
0 \leq \gamma  \leq \gamma_+(1/\bar{u}) \\
 h_f(\gamma, u_f) \leq 0, \quad  \text{for }  u_f \leq u_f^*   \text{ and }  & 
 0 \leq \gamma  \leq \gamma_+(1/\bar{u}) \\
h_f(\gamma, u_f) \leq 0, \quad  
 \text{for all }  u_f  \text{ and } &  \gamma_+(U/\bar{u}) \leq \gamma  \leq \gamma_{max}  
\end{array} 
\right. 
\label{prop-fh} \\ 
&& \text{ } \nonumber \\ 
&& \text{In the interior of phase 2:}    \quad \forall u_f  \in \mathcal{U}, \nonumber \\  
&&  g_h = 1, \quad h_f = 0.  \label{prop-gh2} 
\end{eqnarray} 
with: 
\begin{eqnarray} 
&& u^*_f (\gamma) = \left\{ 
\begin{array}{ll} \displaystyle 
\bar{u} \ln{\left( \frac{ c_1\gamma+c_2 }{c_1\gamma+c_2 -\gamma ^2} \right)} & \text{ for } 
0  \leq \gamma \leq \gamma_+(1/\bar{u}) \\ 
1 & \text{ for  }  \gamma_+(1/\bar{u})  \leq \gamma \leq \gamma_{max} 
 \end{array} 
\right. 
 \label{uf*}  \\ 
&& \gamma_{\pm}(\nu) = \displaystyle \frac{c_1(\nu) \pm \sqrt{c_1(\nu)^2 + 4c_2(\nu)}}{2}, \text{ with } 
c_i(\nu) = c_i(1 - \exp(-\nu)), \quad i = 1, 2.   \label{g+} 
\end{eqnarray} 

\noindent We just sketch the proof (easy but tedious) of these properties. We can first remark that, in the interior of phases 1 and 3 excluding the strip defined by $\gamma_{s_{\pm}}$, 
the $\omega_{s_{\pm}}$ functions  can only take as values $0$ or $1$. In this interior subset $h_f(\gamma,u_f)$ is defined as
\begin{eqnarray*}
h_f(\gamma,u_f) &=& \tau_{hf}(-\gamma ^2 + (c_1\gamma+c_2)(1- \omega_{s}(\gamma) \exp(-u_f/\bar{u})))\\
& =& \tau_{hf} (\gamma_+(u_f/\bar{u}) - \gamma)(\gamma - \gamma_-(u_f/\bar{u}))
\end{eqnarray*}
\noindent It is easy to check that $h_f$ has the sign of $\left(\gamma_+(u_f/\bar{u}) - \gamma\right)$. Besides, $\gamma_{+}$ is an increasing function of $u_f$, since
$$ 
\gamma_+'(\nu) = \frac{c_1\gamma_+(\nu) + c_2 }{\sqrt{c_1(\nu)^2 + 4c_2(\nu)}}\exp(-\nu) 
$$ 
\noindent Hence 0 $\leq \gamma_+(u_f/\bar{u}) \leq \gamma_+(u_{max}/\bar{u}) \equiv \gamma_+(1/\bar{u})$,  so that we can solve $h_f(\gamma^*,u_f) = 0$ and define  $u_f^*(\gamma)$ such that $\gamma_+(u_f^*(\gamma) / \bar{u}) = \gamma $ for $ 0 \leq \gamma \leq \gamma_+(1/\bar{u}) $. Finally, as $\exp{\left(\displaystyle{-\frac{1}{\bar{u}}}\right)}$ is very small,
$$\gamma_+(1/\bar{u})  \approx
(1 - \exp(-1/\bar{u})) \frac{c_1 + \sqrt{c_1^2 + 4c_2}}{2}  < \gamma_{max}$$
This yields property (\ref{prop-fh}). The remaining properties (\ref{prop-gh}) and (\ref{prop-gh2}) are obvious. \\

\noindent Such results imply the following properties of the reachable targets: \\ 
\begin{enumerate}
\item As $g_h >0$, the ages of cells in the reachable targets set have to be larger than those of the cells in their initial conditions. Hence, the reachable targets are on the right of the cell initial positions in the age-maturity plane. 
\item For a given global control $U$ and maturities $0 \leq \gamma  \leq \gamma_+(U/\bar{u})$, property (\ref{prop-fh}) shows that it is possible, using a local feedback $u_f$ around the feedback $u_f^*$, to either increase, decrease or maintain the maturity. In particular, if a target maturity is reached and $U$ remains large enough, the cells remains inside the target. For such targets $\mathcal{M}$, $\mathcal{G}(t ~; t_1,\mathcal{M}) = W(t ~; t_1,\mathcal{M})$. It is also possible to reach a target and then leave it, e.g. if $U$ is lowered. This set-equality 
depends then strongly on the control strategies and positions of the targets with respect to 
$\gamma_+(1/\bar{u})$. 
\item As $U \leq 1$, maturities $\gamma_+(1/\bar{u})  \leq \gamma \leq \gamma_{max}$ are not reachable. 
\end{enumerate}
\medskip 
\noindent Subsequently, we assume that:
\begin{eqnarray} 
&&\text{the target ages are large enough (e.g. $> a_1 + a_2$) } \label{target-a}\\ 
&&\text{the target maturities are below $\gamma_+(1/\bar{u})$ } \label{target-m}
\end{eqnarray}
so that it is not impossible to reach them (but success is not guaranteed, the density conditions needing also to be met). 

\subsubsection{The Hamiltonian in each cellular phase}

\noindent The HamiltonÐJacobi-Bellman (HJB) equations associated with the backwards
reachable sets are solved once a target has been defined and the Hamiltonians in each phase have been 
minimized according to the control terms.  We do not use here the feedforward term: $U_0 = 0$. Hence both the global and local FSH signals operate as best-case feedback controls subject to the explicit constraints 
(rewriting (\ref{normal-u})) $0 \leq u_f \leq U \leq 1$. We note $u = (u_f, U)^T$ the control and $\mathcal{U}$  the set of admissible controls. Such constraints take into account the following physiological knowledge: \\  
(i) the level of FSH in the antral\footnote{ovarian follicles are spheroidal structures hollowed by a cavity called the antrum} fluid is at much as high as FSH plasmatic levels, which implies $u_f\leq U$; \\ 
(ii) there is a basal secretion (i.e. not subject to ovarian feedback) of FSH by the pituitary gland, which implies $U\geq 0$; \\  
(iii) the balance between the secretion rate (or exogenous administration rate) and the clearance rate ensure the saturation of FSH plasmatic levels, which implies $U\leq 1$.

\noindent Two coupled biological questions underlie this reachability problem: (1) which patterns of exogenous FSH administration can be applied to target either ovulation or atresia and (2) how follicular vascularization should develop to remain compatible with those patterns. The answer to such questions might help improving the current FSH treatments, which suffer from several drawbacks such as ovarian overstimulation syndrome. 

\medskip 

\noindent To reach a given target $\mathcal{M}$ from its current location in phase $\Omega_{i,k}$ \footnote{where $i$ stand for the phase index and $k$ for the number of cell generation elapsed since initial time}, a cell (or the lineage of its daughter cells) has to reach some intermediate or final  target in the closure of the $\Omega_{i,k}$ domain. \\ 
For instance, in the case of the ovulatory target $\mathcal{M} = \mathcal{M}_o$ ($\mathcal{M}_o \subset \Omega_{3,N} $) $\mathcal{W}[t_1,\mathcal{M}_o]$ following the solvability tube backward from Phase 3 leads to the boundary between $\Omega_{3,k} $ and $\Omega_{1,k}$ ($k\leq N$); the intersection of 
this boundary with the solvability tube defines a new target set $\mathcal{M}_{o, 1,k}$, in $\Omega_{1,k}$. \\ 
Such intersections depend smoothly upon the width of the thin strip
$0 < \gamma_{s_-} < \gamma < \gamma_{s_+}$ containing these boundaries where the control has little effects. In the following, we suppose that $\gamma_{s_-} = \gamma = \gamma_{s_+}$ so that $\omega_{s_-}$ and $\omega_{s_+}$ are replaced by perfect switch functions. 

\medskip 

\noindent In each cellular phase $i$, in order to compute the backward reachable set 
$\mathcal{G}(t_1 - t ~; t_1,\mathcal{M}_i)$, i.e. the set of states from which it is possible to reach 
the target $\mathcal{M}_i$ in less than a given time $t$, we have to solve an HJB equation of the form 
(\ref{V-intro}) and use the definition (\ref{G-intro}): 
\begin{eqnarray}
V_t + \min (0, H_{fi}(D_xV,x)) &=& 0, \quad t \leq t_1, \quad V(t_1,x) = d^2(x(t_1),\mathcal{M}_i) \label{V}\\ 
\mathcal{G}(t ~; t_1,\mathcal{M}) &=& \{x \in R^n \; | V(t,x) \leq 0\} 
\end{eqnarray}
where $H_{fi}(p,x) = \displaystyle \min_{u \in \mathcal{U}} (p^T.f_i(x,u)) $,  with $p=D_xV$ $=(p_{a_1},...,p_{a_n},p_{\gamma_1},...,p_{\gamma_n},p_{\phi_{f1}},...,p_{\phi_{fn}})^T$. 

\medskip 

\noindent If the targets fulfil the conditions (\ref{target-a}) and (\ref{target-m}) discussed above, we can conclude that: \\ 
i) Starting from cell ages "younger"  than the target ages, for all $u \in \mathcal{U}$, the cells get closer in age 
to target: \\ 
For all $i$, $p_{a_i}.g_f(\gamma_{i}, u_f) \leq 0$, as $p_{a_i} \leq 0$ and $g_f > 0$. \\ 
ii) For the particular choice $U = 1$ and $u_f = u_f^*$, we have $ h_f(\gamma_{i}, u_f) = 0$, so that: \\ 
For all $i$, $p_{\gamma_i}.h_f(\gamma_{i}, u_f^*) = 0$. \\ 
In particular, $ p^T.f_i(x,u_f^*)) \leq 0 $, so that $H_{fi}(p,x) \leq 0$. We have thus shown that in our case, 
solving equation (\ref{V}) amounts to solve the following simpler HJB equation to compute $\mathcal{G}$: 
\begin{eqnarray}\label{Vbis}
V_t + H_{fi}(D_xV,x) &=& 0, \quad t \leq t_1, \quad V(t_1,x) = d^2(x(t_1),\mathcal{M}_i) \\ 
\mathcal{G}(t ~; t_1,\mathcal{M}) &=& \{x \in R^n \; | V(t,x) \leq 0\} 
\end{eqnarray}

\medskip 

\noindent We now compute the Hamiltonian $ H_{fi}(p,x) $ in each phase.
\subsubsection*{Phases 1 \& 3} 
We first compute $p^T.f_i(x,u)$, for $i = 1, 3$. 
$$ p^T.f_i(x,u) = \sum_{c=1}^n p_{a_c}g_{f}(u_f) + \sum_{c=1}^n p_{\gamma_c} h_f(\gamma_c, u_f) + 
\sum_{c=1}^n p_{\phi_{fc}} \phi_{fc} K(\gamma_c, u_f, U) $$  
which can be rewritten as follows: 
\begin{eqnarray*} 
p^T.f_i(x,u) =  H_0(p, x ) -  A(p, x ) \exp(-u_f/\bar{u}) +  B(p) u_f  + C(p, x ) U 
\end{eqnarray*}  
with (using equations (\ref{vitesseg13}), (\ref{vitesseh13}) and (\ref{expressionK})) 
\begin{eqnarray*} 
&& H_0(p, x ) = \tau_{gf} (1 - g_1) \sum_{c=1}^n p_{a_c}  +  \tau_{hf} \sum_{c=1}^n p_{\gamma_c} (-\gamma_c ^2 + c_1\gamma_c + c_2 ) +  \sum_{c=1}^n p_{\phi_{fc}} \phi_{fc} K_0(\gamma_c), \\   
&& \text{where } K_0(\gamma) = - (\omega_{\lambda}(\gamma) + \tau_{hf}(- 2 \gamma + c_1)), \\ 
&& A(p, x ) =  \tau_{hf} \sum_{c=1}^n \left(p_{\gamma_c}(c_1\gamma_c + c_2) - c_1  p_{\phi_{fc}} \phi_{fc} \right), \\Ê 
&& B(p) =  \tau_{gf} g_1\sum_{c=1}^n p_{a_c}, \quad  
C(p, x )  =  \sum_{c=1}^n p_{\phi_{fc}} \phi_{fc} \omega_{\lambda}(\gamma_c).   
\end{eqnarray*}  
Now we minimize this expression with respect to $u \in \mathcal{U}$.  
$$ H_{fi}(p,x) = \displaystyle \min_{u \in \mathcal{U}} (p^T.f_i(x,u)) = H_0(p, x ) +  
\displaystyle \min_{0 \leq u_f \leq U \leq 1}(B(p) u_f -  A(p,x) \exp(-u_f/\bar{u}) + C(p,x) U). $$ 
It is equivalent to minimize first over $U$ with $ u_f \leq U \leq 1$ and then over $u_f$ with 
$0 \leq u_f \leq 1 $ (we drop the arguments $x,p$ for simplicity):  
$$ H_{fi} = H_0  +  \displaystyle \min_{0 \leq u_f  \leq 1}\left( \displaystyle \min_{u_f \leq U \leq 1} \left( B u_f -  
A \exp(-u_f/\bar{u}) + CU \right) \right) $$  
When $C \geq 0$, we have  
$ H_{fi} = H_0 +  \displaystyle \min_{0 \leq u_f  \leq 1}\left( ( B + C) u_f -  A \exp(-u_f/\bar{u}) \right) $. \\Ê
When $C < 0$, we have  
$ H_{fi} = H_0 +  C + \displaystyle \min_{0 \leq u_f  \leq 1}\left( B u_f -  A \exp(-u_f/\bar{u}) \right) $. \\ 
We have then to solve $\displaystyle \min_{0 \leq u_f  \leq 1}\left( \widetilde{B} u_f -  A \exp(-u_f/\bar{u}) \right)$ with 
$\widetilde{B} = B + |C|_+$. \\ 
When $A \geq 0$ (resp. $A < 0$), $ \widetilde{B} u_f -  A \exp(-u_f/\bar{u})$ is a concave (resp. convex) function 
of $u_f$ and the minimizing control is now easy to compute, leading to the following feedback law: 
\begin{eqnarray} 
&& \text{When $A(p,x) \geq 0$: } u_f = 0 \text{  if } -A(p,x) \leq \widetilde{B}(p,x) -  A(p,x) \exp(-1/\bar{u}), \; \text{  else } u_f = 1 \label{A} \\ 
&&  \text{When $A(p,x) < 0$: } u_f =  -\bar{u} \ln (- \frac{\widetilde{B}(p,x)\bar{u}}{A(p,x)})  \text{  if } 
\frac{\widetilde{B}(p,x)\bar{u}}{|A(p,x)|}  \geq \exp(-1/\bar{u}), \text{ else } u_f = 1
\end{eqnarray}   

\subsubsection*{Phase 2}
\noindent This phase is not controlled so that the expression of the
Hamiltonian is straightforward using (\ref{f3}):
\begin{eqnarray*}
H_{f2} &=& \tau_{gf} \sum_{c=1}^n {p_{a_{c}}} + \ln(2)\sum_{c=1}^n {p_{\phi_{fc}}\delta(a_c)\phi_{fc}} 
\end{eqnarray*}

\noindent This analytical minimization enables to write the expression of the Hamiltonian
in each cellular phase.
\subsection{Numerical results}
\label{num_results}
\subsubsection{Implementation: level set methods}
\noindent Two main approaches are used to find reachable sets: the Eulerian
approach approximates the solution values on a fixed grid, whereas the
Lagrangian approach tracks the trajectories of the dynamics. In case of
backwards reachable sets, Eulerian methods are the most appropriate, as they
enable to correctly compute the solution beyond shocks \cite{mitchell}.\\
Amongst the various Eulerian methods available to compute the exact reachable sets, we used
``level set methods'' to solve Eq.(\ref{hamilton}). 
They enable to simulate the motion of dynamic surfaces, such as the zero-level set of
HJB equations \cite{toolboxmitch:04}, with a very high accuracy (about a tenth of the spacing between
the grid points \cite{mitchell}).\\
Such level set algorithms are implemented in the ``Toolbox of Level Set Methods''
\footnote{http://www.cs.ubc.ca/\~{}mitchell/ToolboxLS/}, which can be used to solve, among others, equations of the form:
\begin{eqnarray*}
D_tV(x,t)&+&H(x,D_x V)=0\\
\mbox{with the constraints } 
D_tV(x,t) &\geq& 0 \mbox{ or} \quad V(x,t)\geq \tilde V(x,t)
\end{eqnarray*}
The time derivative, $D_tV$, is approximated by a Runge-Kutta
scheme, with customizable accuracy. The spatial derivative, $D_xV$, is
approximated with an upwind finite difference scheme. The Hamiltonian,
$H(x,p)$, is approximated with a Lax-Friedrichs scheme: 
\begin{eqnarray*}
\hat H(x,p^+,p^-) \equiv H(x,\frac{p^++p^-}{2})-\frac{1}{2}\alpha ^T(p^+-p^-) 
\end{eqnarray*}
where $p^+$ and $p^-$ are respectively the right and left approximations of
$p$, and  $\alpha$ is an artificial dissipation coefficient added to avoid
oscillations in the solution. 

\medskip 

\noindent Although the toolbox is designed for solving initial value problems, it is
possible, in case of autonomous systems, to solve final value problems by
reversing the time in system (\ref{syst}) \cite{toolboxmitch:04}.

\subsubsection{Simulation results}
\begin{figure}[ht]
\begin{center} 
\includegraphics[width=8cm]{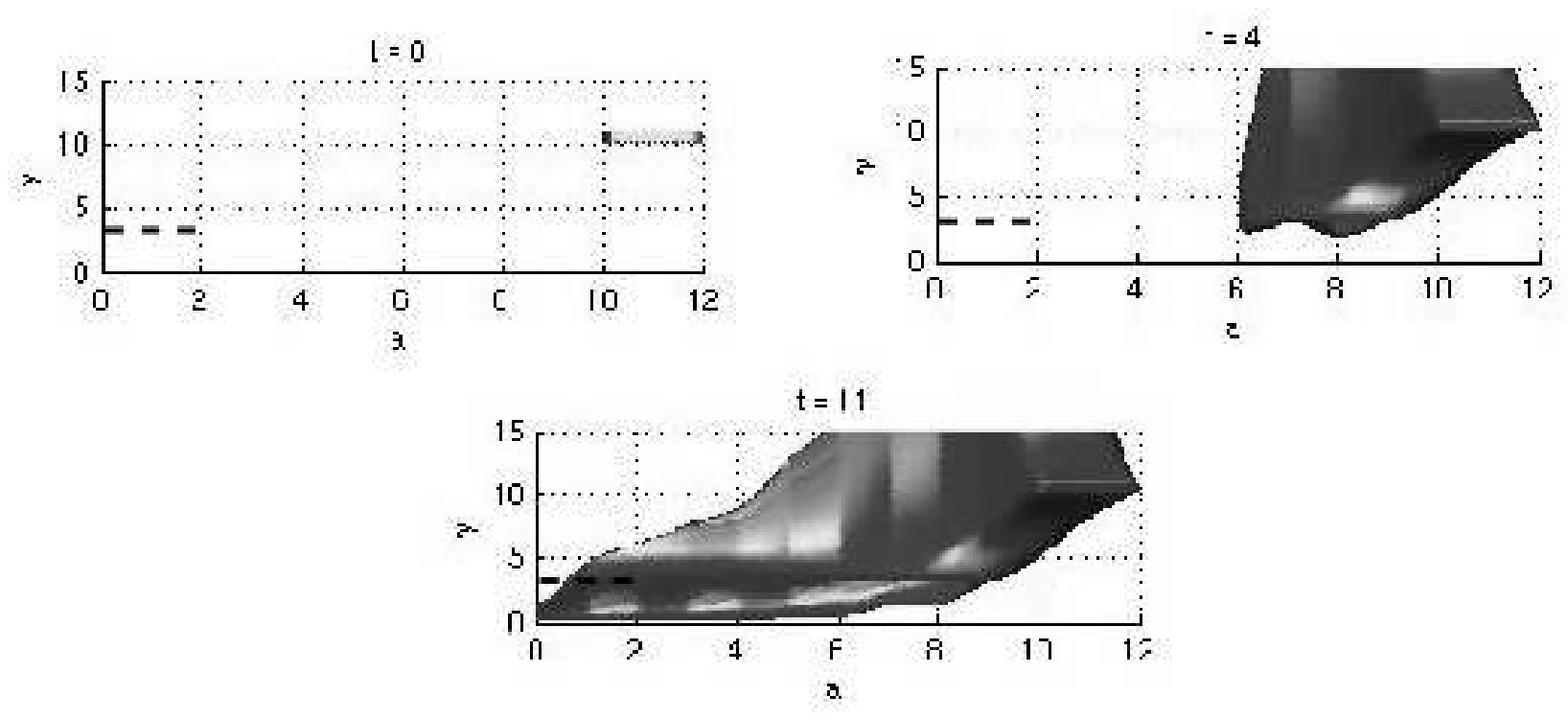}
\includegraphics[width=8cm]{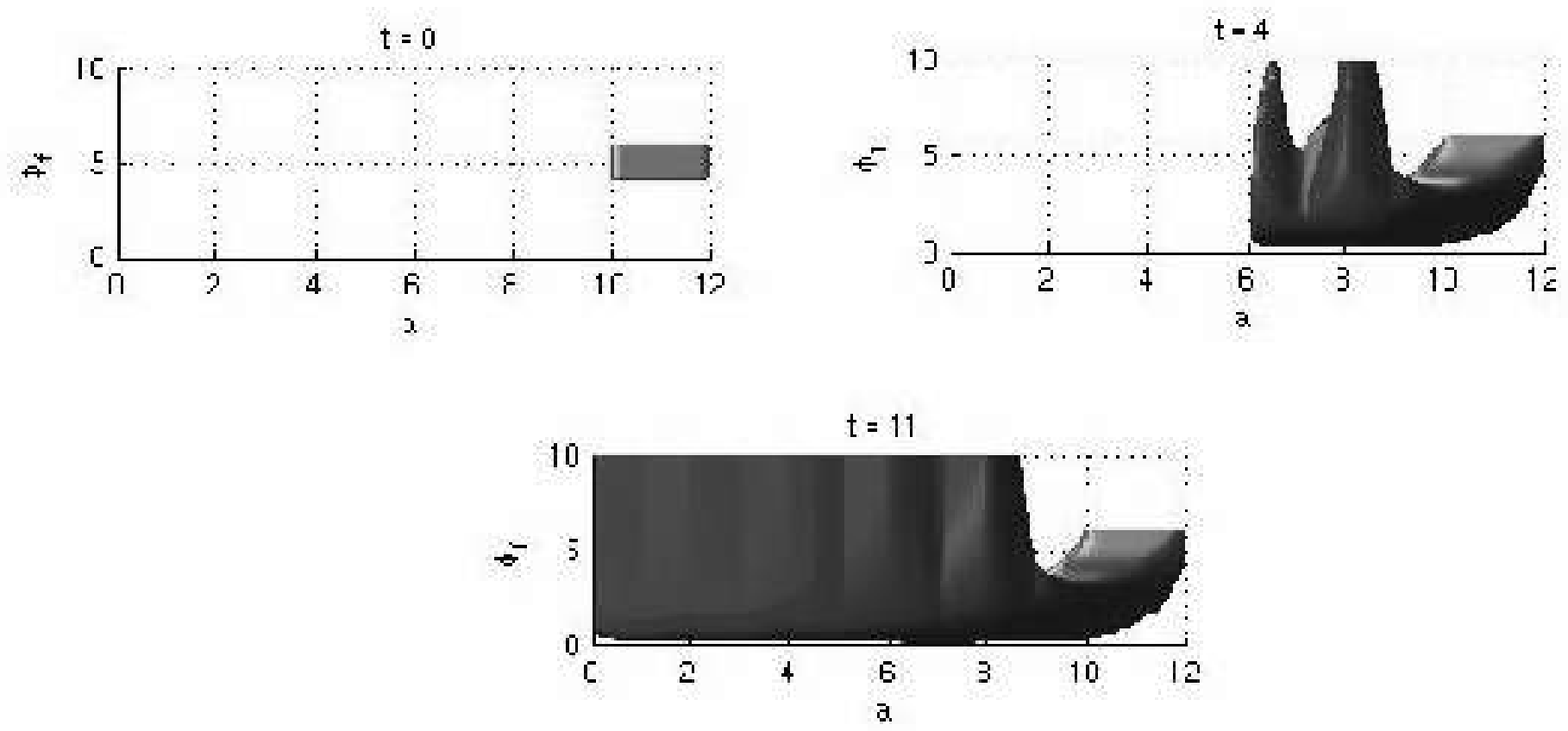}
\caption{Solvability tubes for ovulatory
follicles. The left panel shows the projection of the 3D set on the ($a_1$-$\gamma_1
$) plane, and the right one on the ($a_1$-$\phi_{f1}$) plane.}
\label{ovulation}
\end{center}
\end{figure}

\noindent The initial problem is a $3\times n$ dimensional problem, where $n$ is the number of
characteristics considered. This problem is not tractable from a numerical
ground. Indeed, for $n \geq 2$ and a correct number of grid points
in each dimension ($\simeq 100$) we cope with too many points ($100^{3n}$), and face the dimensionality limits of the algorithms. We thus computed the backwards reachable sets for one characteristic,
$(a_1,\gamma_1,\phi_{f1})$, corresponding to an ovulatory or atretic
trajectory. It is a simplified case where all granulosa cells in a follicle are assumed to be synchronized in age and maturity. \\
In the case of ovulatory follicles, the boundaries of the target
set $\mathcal M_o$ are chosen to coincide with the ranges of age,
maturity, and cell density reached by the characteristics of an ovulatory
follicle at ovulation time, (see Figure \ref{repartition}, left).  \\
\noindent The target set $\mathcal M_o$ for ovulation is thus defined as:
\begin{equation}
\mathcal M_o =\{(a_1,\gamma_1 ,\phi_{f1} )\in [10,12]\times [10,11]\times [4,6]\}
\end{equation}
\noindent Figure \ref{ovulation} illustrates the trajectories compatible with ovulation.\\ 
The left panel illustrates the projection, on the
age-maturity plane, of the changes in the backwards reachable set. Time $t=0$
(first subplot) is ovulation time, and we can see the rectangle $\{[10,12]\times [10,11]\}$
representing the
projection of the target set $\mathcal M_o$ (the age range is slightly different from that of Figure \ref{repartition} due to difference in the dimensioning of the unrolled domain compared to the periodic domain). At time $t=4$, the set area has
increased, including all states that can join the target set in
less than 4 time units (2 time units roughly correspond to 1 cell cycle duration). 
The backwards reachable set at time $t=11$ represents the set of states that can reach the target set for ovulation in at most 11 time units. We chose to stop the simulation at this time as it contains the
initial conditions used for the simulation represented in Figure \ref{repartition}.
The ``staircase'' shape of the backwards reachable set is due to the dynamics
in the cell cycle. The maturity increases with age in phase 1, while its remains unchanged in phase 2, yielding to the flat part of the staircase. An instance of the flat
part can be seen at time $t=11$ between $a_1=7$ and $a_1=8$.\\
The right panel illustrates the projection, on the
age-cell density plane, of the changes in the backwards reachable set. We can again see the
projection  of the target set $\mathcal M_o$ at time $t=0$, as the rectangle $\{[10,11]\times [4,6]\}$. At time $t=11$, a large
part of the age-cell density domain is compatible with reaching the target set.
From age $a_1=9$ onward, the cell density variable increases with age to reach the target set. This is in agreement with the cell density dynamics in phase 3.  

\medskip 

\noindent The backwards reachable set of $\mathcal M_o$ nearly contains
the whole spatial domain. It is an overapproximation of the initial
states that allow to reach the target set. Indeed, it does not only contain the
trajectories of the characteristics of the ovulatory follicle, but
also those starting from initial conditions that do not arise on a
physiological ground. Basing on biological
knowledge, we can thus select from the overapproximated set, the
physiologically-meaningful subset: we assume that all cells are initially within the cell cycle, at
their first generation, so that the subset of admissible initial conditions is defined by
$\{(a_1(t_0),\gamma_1(t_0))\in [0,a_2]\times [0,\gamma_s]\}$. 
This subset is below the dashed segment on the left panel of Figure \ref{ovulation}. 

\medskip 

\noindent The target set $\mathcal M_a$ for atresia is defined by the ranges reached by
the characteristics of atretic follicles:
\begin{equation}
\mathcal M_a =\{(a_1,\gamma_1 ,\phi_{f1} )\in [8,10]\times [3,4]\times [2,4]\}
\end{equation}

\begin{figure}[ht]
\begin{center}
\includegraphics[width=8cm]{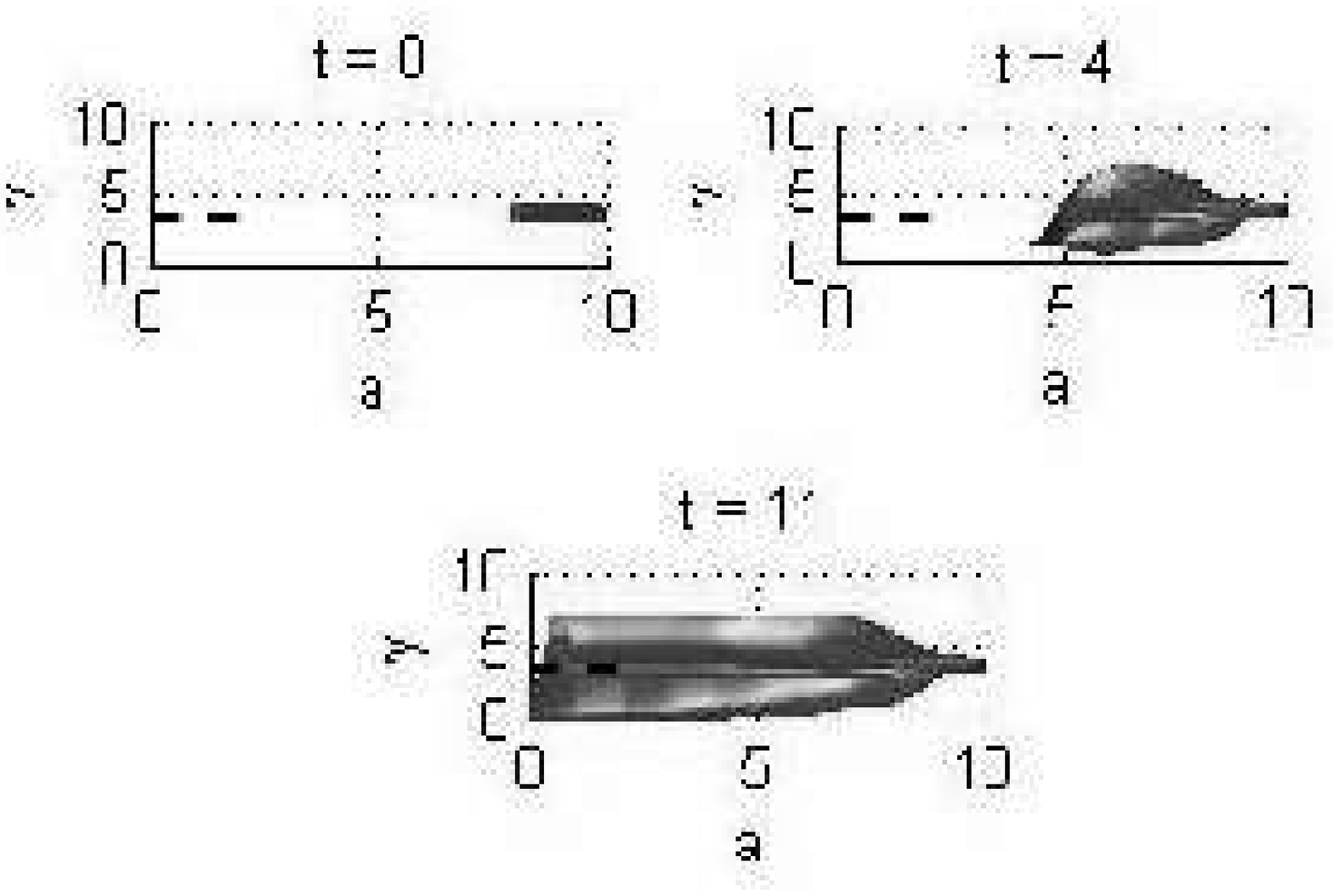}
\includegraphics[width=8cm]{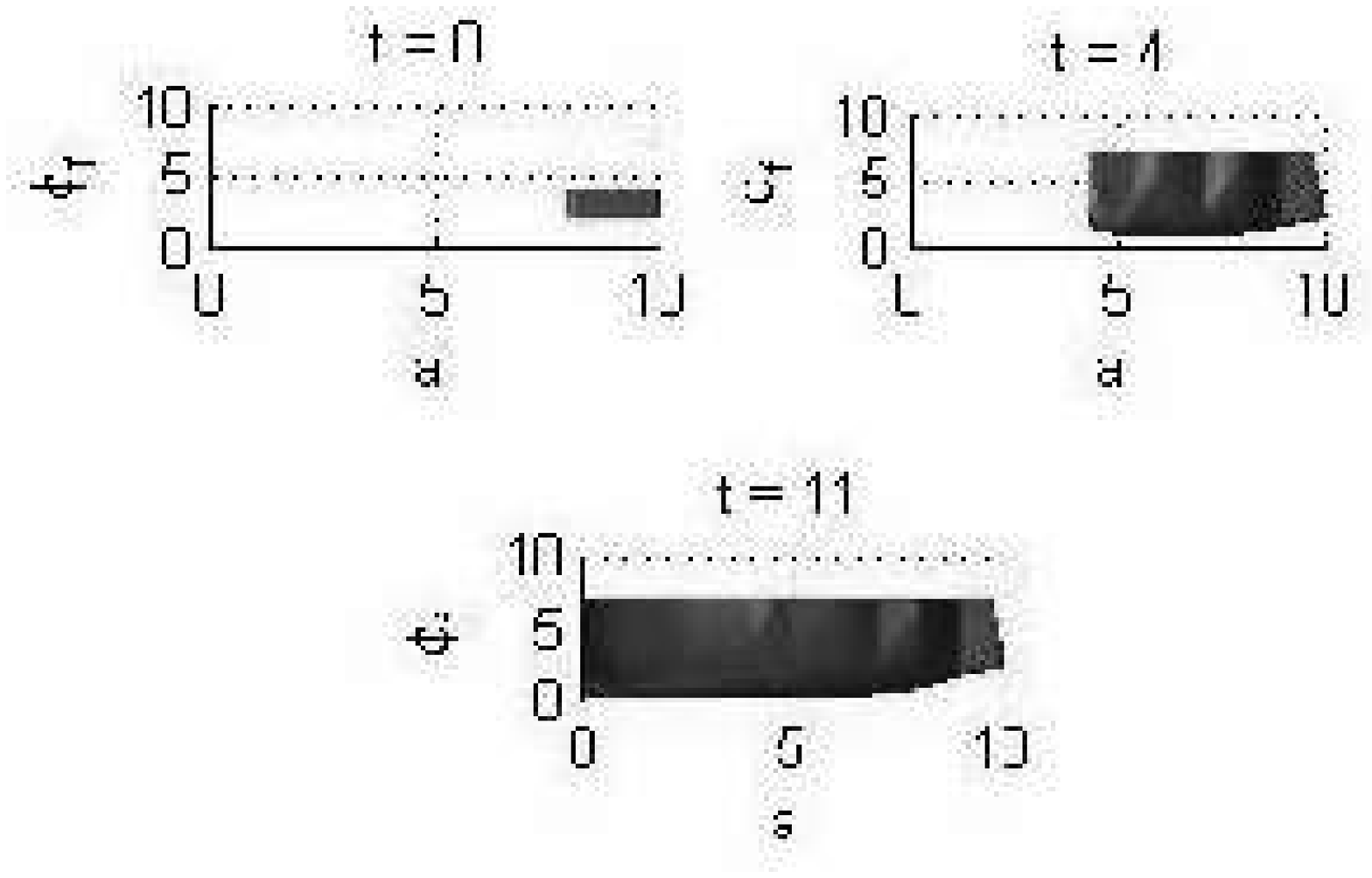}
\caption{Solvability tubes for atretic follicles. The left panel shows the projection of the 3D set on the ($a_1$-$\gamma_1
$) plane, and the right one on the ($a_1$-$\phi_{f1}$) plane.}
\label{atresie}
\end{center}
\end{figure}

\noindent Figure \ref{atresie} illustrates all the trajectories that fall into the target
set for an atretic follicle. The left panel illustrates the projection, on the
age-maturity plane, of the changes in the backwards reachable set. Some states in the target set are initially in the cell cycle, leading to the ``staircase'' pattern described below. The right panel illustrates the projection, on the
age-cell density plane, of the temporal changes in the backwards reachable set. The set area
increases, and at time $t=11$, it roughly contains the whole domain. Once again, the set of initial states leading to atresia is overapproximated, and we can apply the same constraints as in the ovulatory case on the initial conditions, that are again below the dashed segment on Figure \ref{atresie}.

\noindent The initial conditions leading to ovulation or atresia, subject to physiological constraints, are roughly
superimposed, and contain the whole subset of admissible initial conditions. Thus we cannot distinguish {\it a priori} ovulatory
trajectories from atretic ones. This result agrees with physiological
knowledge, as there is no predestination in follicular fate
\cite{gougeon_96}.\\
Although the initial conditions do not enable to determine whether a follicle
is ovulatory or atretic, there is a moment when trajectories leading to either ovulation
or atresia diverge. Such a moment corresponds to the
physiological time of selection, when the trajectories of the few ovulatory follicles separate from the atretic ones.
\section{Discussion and perspectives}
\noindent We have characterized the backwards reachable sets for both ovulatory and
atretic follicles. This control problem has been solved by studying the
characteristics of the conservation laws describing the dynamics of the structured granulosa cell populations in ovarian follicles. The HJB equations representing the backwards reachable sets
have been solved numerically for one characteristic curve, and allow to characterize the set of initial conditions that lead a
characteristic into the target set either for ovulation or atresia. 

\medskip 

\noindent We have used the backwards reachable sets theory for a problem that is not
entirely continuous, as there are discontinuities at the transitions
between each cellular phase. Hence, we studied each phase separately,
assuming that the transitions at the boundaries were
continuous. Actually, we deal with a hybrid problem, as each cellular phase has
its proper dynamics. As far as time dependent HJB equations are
concerned, elements of a theory for hybrid systems are established, especially concerning ``reach-avoid'' sets \cite{tomlin}, but it is not complete
yet, and not directly applicable to our problem. Viability theory also considers reachable sets for hybrid systems \cite{aubin}, but the
algorithms used are not as accurate as level set methods, since they use a
different representation of reachable sets \cite{mitchell}. Yet the numerical
results we have obtained with the level set methods seem correct since the continuous
part of the system is solved with highly accurate algorithms and the hybrid
part is handled via continuity conditions and substantiated by the trace results obtained on the conservation laws. 

\medskip 

\noindent We have solved a simplified numerical problem dealing with a single
characteristic of a follicle, as the simulation tool needs too huge memory to solve
higher dimensional problems. The backwards reachable sets
obtained show that it is possible to find
a correct control law to steer a large set of initial conditions into the
target set, and they give information upon the maximum duration needed to reach the
target.\\
If the simulation of more than one characteristic were possible, we expect that the backwards reachable
sets would be smaller. We even speculate that, from a physiological viewpoint, a larger set in case of a single growing
follicle ensures the ovulation success, while a smaller set in case of several
simultaneously growing follicles limits the ovulation rate and underlies the selection process  in part.
Yet this assumption needs improvement of the numerical implementation to be tested. 

\medskip 

\noindent We have studied by now the open-loop problem, for the ovulation of one
follicle. This situation is close to therapeutic situations of ovarian
stimulations, so that the designed control laws may be useful in improving
therapeutic schemes. Yet, to understand ovulation control in depth, we will have to tackle multi-scale, closed-loop reachability problems of selection, which is the matter of future work.

\bibliographystyle{unsrt}
\bibliography{bibbrs-2}
\end{document}